\documentclass[useAMS,usenatbib]{mn2e}

\usepackage{graphicx}
\usepackage{url}
\usepackage{amsmath}

\def\apgt{\ {\raise-.5ex\hbox{$\buildrel>\over\sim$}}\ }
\def\aplt{\ {\raise-.5ex\hbox{$\buildrel<\over\sim$}}\ }
\def\lteq{\ {\raise-.5ex\hbox{$\buildrel<\over-$}}\ }

\title[How Sedna was captured from a solar sibling]
{How Sedna and family were captured in a close encounter with a solar sibling}

\author[L. J\'{i}lkov\'{a} et al.]
{Lucie J\'{i}lkov\'{a},$^{1}$\thanks{E-mail:jilkova@strw.leidenuniv.nl\,(LJ); spz@strw.leidenuniv.nl\,(SPZ)}
Simon Portegies Zwart,$^{1}$\footnotemark[1]
Tjibaria Pijloo$^{1,2}$ and 
Michael Hammer$^{3}$
\\
$^{1}$ Leiden Observatory, Leiden University, PO Box 9513, 2300 RA, Leiden, The Netherlands \\
$^{2}$ Department of Astrophysics/IMAPP, Radboud University, PO Box 9010, 6500 GL Nijmegen, the Netherlands \\
$^{3}$ Cornell University, 614 Space Sciences Building, Ithaca, NY 14853
}

\date{Accepted . Received ; in original form 2015 March 9}
\pagerange{\pageref{firstpage}--\pageref{lastpage}} \pubyear{2002}

\begin{document}

\label{firstpage}

\maketitle

\begin{abstract} 
  The discovery of 2012\,VP$_{113}$ initiated the debate on the origin of the Sedna family of planetesimals in orbit around the Sun.
  Sednitos roam the outer regions of the Solar system between the Egeworth--Kuiper belt and the Oort cloud, in extraordinary wide ($a >150$\,au) orbits with a large perihelion distance of $q > 30$\,au compared to the Earth's ($a \equiv 1$\,au and eccentricity $e \equiv (1-q/a) \simeq 0.0167$ or $q\simeq 1$\,au). 
  This population is composed of a dozen objects, which we consider a family because they have similar perihelion distance and inclination with respect to the ecliptic $i = 10^\circ$--$30^\circ$.
  They also have similar argument of perihelion $\omega = 340^\circ \pm 55^\circ$. 
  There is no ready explanation for their origin. 
  Here we show that these orbital parameters are typical for a captured population from the planetesimal disk of another star. 
  Assuming the orbital elements of Sednitos have not changed since they acquired their orbits, we reconstruct the encounter that led to their capture.
  We conclude that they might have been captured in a near miss with a $1.8$\,$\rm{M}_{\sun}$\, star that impacted the Sun at $\simeq 340$\,au at an inclination with respect to the ecliptic of 17--$34^\circ$ with a relative velocity at infinity of $\sim 4.3$\,km/s.
  We predict that the Sednitos-region is populated by 930 planetesimals and the inner Oort cloud acquired $\sim 440$ planetesimals through the same encounter.
\end{abstract}


\section{Introduction}
\label{sec:intro}

Upon its discovery 90377 Sedna \citep{2004ApJ...617..645B} was proposed to originate from the Edgeworth--Kuiper belt; a population of rocks in an almost planar disk aligned with the ecliptic and much closer to the Sun ($q = 30$--50\,au) than the Oort cloud. 
According to this model the violent and rather sudden migration of Uranus and Neptune would have excited the cold Kuiper belt \citep{2012Icar..217....1B}.  
This reorganization of the outer giant planets would have initiated the hot Kuiper belt \citep{2008Icar..196..258L} that, due to the long local
relaxation time, cools down only very slowly \citep{2014MNRAS.444.2808P}. 
Subsequent chaotic diffusion \citep{2008DDA....39.1204M}, perturbations in the Sun's birth environment in close flybys 
\citep{2014prpl.conf..787D},
or more distant encounters could have caused further migration of the Kuiper-belt objects to orbits similar to that of Sedna \citep{2012Icar..217....1B}.

Recently, \citet{2014Natur.507..471T} discovered the object 2012\,VP$_{113}$, a second member of the inner Oort cloud, which they defined as a family of planetesimals with $q\apgt 50$\,au and $a\approx 150$--1500\,au.
They furthermore identified a population of planetesimals between the Edgeworth--Kuiper belt and the Oort cloud that share similar orbital elements \citep[see also][]{2014MNRAS.443L..59D}, namely the large perihelion and semi-major axis ($q>30$\,au and $a>150$\,au, respectively), inclination with respect to the ecliptic ($i=$10--30$^\circ$), and the argument of perihelion ($\omega=340^\circ \pm 55^\circ$).
Currently 13 such objects have been observed in the outer Solar system and it was suggested that their characteristics resulted from a common origin \citep{2014Natur.507..471T,2014MNRAS.443L..59D}.
Here we consider this group of object a family which we call {\it Sednitos}.

The perihelion distances of Sedna and 2012\,VP$_{113}$ are too large for any Kuiper belt object and their aphelion distances are too short for them to be Oort cloud objects \citep{2015MNRAS.446.3788B}.  
It is therefore hard to explain them as members of either population.
In principle chaotic diffusion could cause sufficient internal migration, but at the distance of Sedna the time scale for this process exceeds the age of the Solar system \citep{1988Sci...241..433S}.
If Sedna stood alone, such an exotic explanation could be satisfactory.
However, this cannot explain the entire population of the inner Oort cloud which, when taking selection effects into account, amounts to $430^{+400}_{-240}$ members brighter than $r=$24.3\,mag \citep{2014Natur.507..471T}.
In the model of \citet{2012Icar..217....1B} the inner Oort objects are scattered from the Kuiper belt and decoupled to larger pericenters by perturbations in the Sun's birth cluster. The size of the population produced by this mechanism is consistent with the one predicted from observations.

According to an alternative scenario, Sedna could have been captured from the outer disk of a passing star, as suggested by \citet{2004AJ....128.2564M} and \citet{2004Natur.432..598K}.
They showed that capturing a planetesimal into a Sedna-like orbit is possible, but they did not carry out a detailed parameter space study.
The model of \citet{2004Natur.432..598K} could account for at most $10$\,\% of the Sednitos and it was tuned at producing the outer edge of the Edgeworth--Kuiper belt at the currently observed 50\,au; however, this is inconsistent with the {\em Nice} model that requires an edge at $\sim 35$\,au \citep{2004Icar..170..492G}.
The capture of planetesimals by the Solar system was further studied by \citet{2010Sci...329..187L}, who simulated the Sun's birth cluster considering the transfer of planetesimals among stars.
However, the study was aimed to explain the origin of the Oort cloud and the orbits of most of the captured objects have large semi-major axes ($a\apgt10^3$\,au) and perihelia ($q\apgt10^2$\,au), not representative for Sedna and 2012\,VP$_{113}$.

\subsection{Argument of perihelion of Sednitos}
\label{sec:omega}
Sednitos are characterized by a clustered distribution of their observed argument of perihelion $\omega$.
The precession period of $\omega$ depends on the semi-major axis, eccentricity, and inclination of the precessing orbit.
The precession periods for all Sednitos excluding Sedna range from about 40\,Myr up to 650\,Myr, while Sedna has the longest precession period of about 1.5\,Gyr \citep{2014Natur.507..471T,2006Icar..184...59B,2006Icar..184..589G}.
Therefore the clustering of $\omega$ must have happened relatively recently (less than few Myr ago) or a dynamical mechanism must have been constraining the distribution of $\omega$ since it was established.
\citet{2014Natur.507..471T} suggested that an outer Solar system pertuber of 5--15\,M$_{\rm{Earth}}$ orbiting the Sun between 200 and 300\,au is restricting the Sednitos' evolution in $\omega$ by the Kozai--Lidov mechanism \citep{kozai_secular_1962,lidov_evolution_1962}.
Based on further analysis of the Sednitos' orbital elements, \citet{2014MNRAS.443L..59D} suggested that at least two planetary-mass trans-Neptunian perturbers exist at approximately 200\,au and 250\,au.

The perturbing object (possibly more than one object) is assumed to be on a low-inclination almost circular orbit.
When, in such a configuration, the ratio of semi-major axis with respect to the perturbed objects (i.e. Sednitos) is close to 1, the relative argument of perihelion of the perturbed and perturbing orbits can librate around $0^{\circ}$ or $180^{\circ}$ due to the Kozai--Lidov mechanism \citep[see the Extended materials of][for an example where $\omega$ of 2012\,VP$_{113}$ librates in the range $0^{\circ}\pm60^{\circ}$]{2014Natur.507..471T}.
However, depending on the initial relative inclination and argument of perihelion of the perturbing and perturbed orbits, the argument of perihelion can also circulate (i.e. periodically change values from $-180^{\circ}$ to $180^{\circ}$).
The libration around $0^{\circ}$ will occur if the initial relative $\omega$ ranges from $-90^{\circ}$ to $90^{\circ}$ \citep[e.g.][]{2007MNRAS.382.1768M}.
Therefore, the $\omega$ of Sednitos relative to the perturber ($\omega-\omega_{\rm{pertuber}}$) needs to be constrained at the beginning of the dynamical interaction with the perturber.
\citet{2006Icar..184...59B} showed that for the Sedna-like orbits produced during the early evolution of the Solar system, when the Sun was still residing in its birth star cluster, preferentially $\omega=0^{\circ}$ or $180^{\circ}$.
This mechanism could therefore explain the initial clustering of Sednitos' argument of perihelion, although it is not clear why the orbits initially obtained $\omega$ about $0^{\circ}$ and not $180^{\circ}$ \citep[see also][]{2014Natur.507..471T}. 

The presence of the perturbing object(s) in the outer Solar system is currently the only mechanism suggested to explain how the clustering in $\omega$ is preserved on timescales longer than the precession periods of Sednitos.
At the same time, \citet{2014MNRAS.444L..78I} ruled out the presence of super-Earth planet of 2--15\,M$_{\rm{Earth}}$ with $a\approx 200$--300\,au using the current constraints on the anomalous secular precession of the argument of perihelion of some of the known planets in the Solar system.
Therefore, the existence of an outer planet is still under debate.

Irrespective of the mechanism that preserves the clustering of the argument of perihelion, we present that such clustering is a general characteristic of the population transferred during a stellar encounter.
The constrained distribution in $\omega$ can then be shepherded by some other process.
We argue that Sednitos are found in the Parking zone of the Solar system where their semi-major axis and eccentricity have been unaffected once the Sun left its birth cluster \citep{2015MNRAS.451..144P}.
We therefore use the current semi-major axis and eccentricity to constrain the encounter that might have introduced the Sednitos into the Solar system.

\section{Methods}
\label{sec:methods}

The encounter between the Sun and another star, here called~Q, with a planetesimal disk can be simulated by integrating the equations of motion of the two stars using a symplectic $N$-body code. We use {\tt Huayno} \citep{2012NewA...17..711P} for this. 
As long as the two stars are well separated (at least three times the disk size) we integrate the planetesimals using {\tt Sakura} \citep{2014MNRAS.440..719G}, in which Kepler's equations are solved in the potential of the Q coupled with the perturbations from the Sun.
The planetesimals are represented by zero-mass particles which do not affect each other and neither the motion of the two stars, while the planetesimals themselves are affected by the two stars.
Both integrators ({\tt Huayno} and {\tt Sakura}) are coupled via {\tt Bridge} \citep{2007PASJ...59.1095F}, which is an extension of the mixed variable symplectic scheme \citep{1991AJ....102.1528W} and is used in this context to couple two different dynamical regimes within one
self-gravitating system.  The coupling of codes is realized using the
Astronomical Multi-purpose Software Environment
\citep{2013CoPhC.184..456P}.\footnote{All source code is available at \url{http://amusecode.org}.}
When the stars move close enough that the planetesimals orbits can no longer be considered Keplerian\,---\,i.e. the motion is no longer dominated by Q and both stars have a substantial influence on the orbits\,---\,the planetesimals are integrated directly using {\tt Huayno}.
We introduce this transition from hybrid to direct integration when the time since the beginning of the encounter equals half of the time for the two stars to reach their closest approach (which always results to a separation of the two stars larger than three times the disk size).

The initial conditions for the encounter are as follows. 
The distance between the two stars is determined by the condition that the magnitude of the gravitational force from the Sun at the outer edge of Q's disk equals 10\% of Q's force.
We tested that increasing the initial separation does not change the results.
In all our calculations we adopted the mass of the Sun of 1\,$\rm{M}_{\sun}$.
Planetesimals in Q's disk have initially planar distribution and their radial distance from Q, $r$, follows a uniform random distribution, i.e. the surface density profile of Q's disk $\propto (1/r)$.
However, since the planetesimals are represented by zero-mass particles, the surface density profile can be adjusted in post-processing (see Sec.~\ref{sec:results}).
The planetesimals are initially on circular orbits.
The inner edge of Q's disk is 10\,au.
We set the upper limit on the outer edge of Q's disk, $r_{\rm max, Q}$, to 200\,au and determine the actual value from the minimal requirement of producing planetesimals in the range of $q=30$\,au to 85\,au; we do the same for Sun's disk, see below.

The encounter between the Sun and Q is characterized by the five parameters 
(also listed in Tab.\,\ref{Tab:Sednitos})\,---\,the mass of the encountering star, $M_{\rm Q}$, the closest approach of the stars, $q_{\rm Q}$, the eccentricity of the orbit, $e_{\rm Q}$, the inclination of the encounter plane with respect to Q's disk, $i_{\rm Q}$, and the argument of periastron of the orbit, $\omega_{\rm Q}$.
We have the computer map this parameter space automatically using the affine-invariant, parallel stretch-move algorithm for Markov Chain Monte Carlo \citep{HASTINGS01041970} with specific optimizations \citep{1189.65014} using {\tt emcee} \citep{2013PASP..125..306F}.

\begin{table}
  \caption{ Reconstructed encounter parameters for the star that delivered the Sednitos into the Solar system.  
    The first column lists the five parameters of the encounter:
    $M_{\rm Q}$ is the mass of the impactor star~Q,
    $q_{\rm Q}$ the closest approach of the Sun and Q,
    $e_{\rm Q}$ the eccentricity of their orbit,
    $i_{\rm Q}$ the inclination of the orbital plane with respect to Q's disk, 
    and $\omega_{\rm Q}$ the argument of periastron.
    We further give the impact parameter $b$, 
    the relative velocity of the encounter at infinity $v$, 
    and the limits for the outer edges for Q's and Sun's disk, $r_{\rm max, Q}$ and $r_{\rm max, \odot}$, respectively. 
    The orientation of the encounter with respect to the ecliptic is specified by $i_{\rm enc}$  and $\omega_{\rm enc}$.
    The second column gives the range considered in the Markov chain simulations, followed by the range of parameters that led to a satisfactory solution. 
    The parameters of the preferred encounter are listed in the right most column (we give the constrained range for $i_{\rm enc}$ and $\omega_{\rm enc}$ together with the individual values used in the presented example in the parenthesis).
  \label{Tab:Sednitos}
}
  \begin{center}
  \begin{tabular}{llll}
  \hline
  parameter             & parameter           & viable          & preferred \\
                        & range               & range           & encounter \\
\hline
  $M_{\rm Q}$           & 0.2--2.0\,$\rm{M}_{\sun}$             & 1.0--2.0\,$\rm{M}_{\sun}$   & 1.8\,$\rm{M}_{\sun}$ \\
  $q_{\rm Q}$           & 200--393\,au       & 210--320\,au     & 227\,au \\
  $e_{\rm Q}$           & 1.001--4.0         & 1.9--3.8         & 2.6 \\
  $i_{\rm Q}$           & 0--180$^\circ$     & 2--44$^\circ$    & 35$^\circ$ \\
  $\omega_{\rm Q}$      & 0--180$^\circ$     & 0--180$^\circ$   & 175$^\circ$ \\
  $b$                   & 265--2071\,au      & 280--450\,au     & 340\,au \\
  $v$                   & 0.4--6.0 km/s      & 3.1--5.4 km/s    & 4.3 km/s \\
  $r_{\rm max, Q}$      &               & 130--200\,au          & $\apgt 161$\,au\\
  $r_{\rm max, \odot}$  &               &                       & $\aplt 70$\,au\\
  $i_{\rm enc}$         & & 0--70$^\circ$  & 17--34$^\circ$ (28$^\circ$) \\
  $\omega_{\rm enc}$    & & 0--360$^\circ$ & 154--197$^\circ$ (170$^\circ$) \\
  \hline
  \end{tabular}
  \end{center}
\end{table}

We run more than 10,000 realizations of possible encounters. 
Each calculation is performed with up to 20,000 particles (in chunks of 500) in the planetesimal disk until the number of particles captured
by the Sun amounts to at least 13 objects with a perihelion distance between 30\,au and 85\,au.
To account for the observability of orbits with different eccentricities, we weight each particle by the time it spends within 85\,au from the Sun measured as a fraction of the orbital period.
The weight $w$ is calculated using the mean anomaly at the 85\,au from the Sun, $M(85\,\mathrm{au})$,
\begin{equation}
w= \begin{cases} 
M(85\,\mathrm{au})/\pi & \mathrm{if\,\,aphelion}>85\,\mathrm{au}, \\ 
1 & \mathrm{if\,\,} 30\,\mathrm{au}<\mathrm{aphelion}<85\,\mathrm{au}. 
\end{cases}
\end{equation}
This weighting favors finding relatively low eccentricity orbits with a small perihelion, as is consistent with how the Sednitos were discovered \citep{2014Natur.507..471T}.

The resulting distribution of the planetesimals in semi-major axis and eccentricity is subsequently compared with the observed dozen Sednitos listed in Table~2 of \cite{2014Natur.507..471T} and Table~1 in \cite{2014MNRAS.443L..59D} for 2003\,SS$_{422}$.
The 13 observed Sednitos provide only low-number statistics and we investigate what constraints can we draw based on the limited data.
As a first step in the statistical comparison, we perform a consistency analysis between the simulated objects and the observed Sednitos under the hypothesis that the latter is a random sub-sample of the former using multivariate analysis. 
We calculate the ranking of the Henze statistic \citep{1988AnnSt...16..772H,2007MNRAS.377.1281K} using the nearest neighbors and based on 500 randomly pooled data \citep[see][for more detail]{2007MNRAS.377.1281K} and we require the final rank (or the $p\mbox{-value}$) of the actual data sets to be $>0.05$ to consider the samples consistent.
In these cases, we measure the separation distance between these two distributions in the plane of $a$ vs. $e$ using the Hellinger distance of binned kernel smoothed distributions \citep{1998AnnSt...50..503P}.
We use a grid of $20\times20$\,bins scaled on the observed data with a symmetric Gaussian kernel with a relative width (corresponding to the standard deviation) of 0.08.
The separation that emerges from this analysis is used as the posterior probability in the Markov Chain Monte Carlo.

\section{Results}
\label{sec:results}

We identify a region in the 5-dimensional encounter parameter space where the $a$ vs. $e$ distribution of the simulated transferred particles is statistically indistinguishable from the observed distribution.
We constrain the viable range by the Hellinger distance $<0.6$.
We further divide the Sednitos $a$ vs. $e$ region into three sections: the inner Oort cloud ($70\,\mathrm{au}<q<85$\,au), the region where so far no objects have been observed ($50\,\mathrm{au}<q<75$\,au, or $q<50\,\mathrm{au}$ and $a<140\,\mathrm{au}$), and the remaining region.
For the encounters in the viable range, we require that the weighted number of captured particles is higher then 1.0 in the inner Oort cloud region, and lower than 6.0 for the region without any observed objects.

The limits of such viable range for individual parameters are listed in Tab.\,\ref{Tab:Sednitos}.
We also select a preferred encounter as a representative example of the viable range;
its parameters are also listed Tab.\,\ref{Tab:Sednitos}.
The Sednitos' distribution produced by the preferred encounter compares well with the observed one: the rank of the Henze statistics is 0.5 and the distance between the distributions is $\sim0.5$ (here 0 corresponds to identical binned kernel smoothed distributions, while 1 corresponds to distributions with no overlap; see Sec.~\ref{sec:methods}).

After the Markov chain calculation we constrain the orientation of Q's disk with respect to the ecliptic.
While the inclination $i_{\rm Q}$ and the argument of periastron $\omega_{\rm Q}$ of the orbital plane of the encounter with respect to Q's disk are constrained by the Markov chain calculations, the orientation of the orbital plane with respect to the ecliptic is unconstrained.
We constrain the orientation using two-sample Kolmogorov--Smirnov tests comparing each of the distributions in the argument of perihelion, inclination, and the longitude of the ascending node of the observed and simulated Sednitos.
The clustering of inclination and argument of perihelion is a general feature of the transferred population and an orientation of the coordinate system where the simulated distributions are consistent with the observed ones is found for almost all viable encounters.
Using a grid with a step size of $2^{\circ}$ for each of the three Euler angles, we rotate the coordinate system centered on the Sun until the $p\mbox{-value}>0.05$ for each of the three compared distributions.
We derive the inclination $i_{\rm enc}$ and the argument of periastron $\omega_{\rm enc}$ of the orbit of the encounter with respect to the ecliptic (the longitude of ascending node is a free parameter due to assumed symmetry of the Sun's disk, see below).
This procedure results in the inclination $i_{\rm enc}$ and the argument of perihelion $\omega_{\rm enc}$ of the encounter with respect to the ecliptic for individual encounters.
These parameters are typically constrained within intervals of $\pm10^\circ$ and $\pm20^\circ$, for $i_{\rm enc}$ and $\omega_{\rm enc}$ respectively.
We summarize the values in Tab.\,\ref{Tab:Sednitos} (note that while always constrained within limited intervals for individual encounters, $\omega_{\rm enc}$ have values in a wide range, unlike $i_{\rm enc}$).

\begin{figure}
\begin{center}
\includegraphics[width=0.45\textwidth,angle=-0.0]{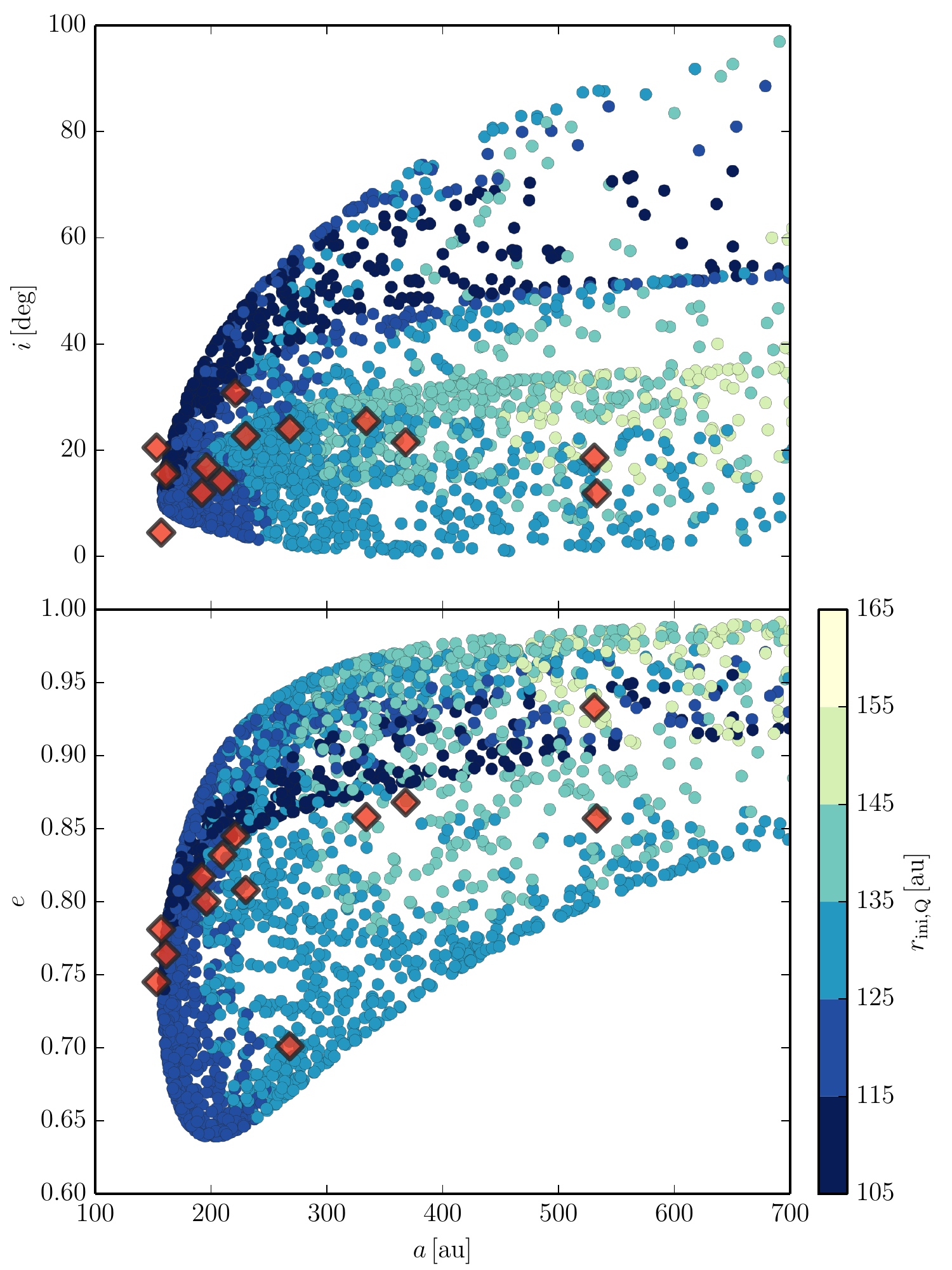}
\end{center}
\caption{Distribution of the planetesimals captured around the Sun during the preferred encounter with star~Q (right most column of Tab.\,\ref{Tab:Sednitos}).
Along the $x$-axis we present the semi-major axis $a$ of the captured planetesimals.
The top panel gives the inclination $i$ along the $y$-axis, and the bottom panel gives the eccentricity $e$.
The color scale maps the initial radius in Q's disk, $r_{\rm ini, Q}$.
Note that the simulated particles are not weighted here.
The red diamonds give the observed positions of the Sednitos.
\label{Fig:Sednitos}}
\end{figure}

\begin{figure*}
\begin{center}
\includegraphics[width=0.4\textwidth,angle=-0.0]{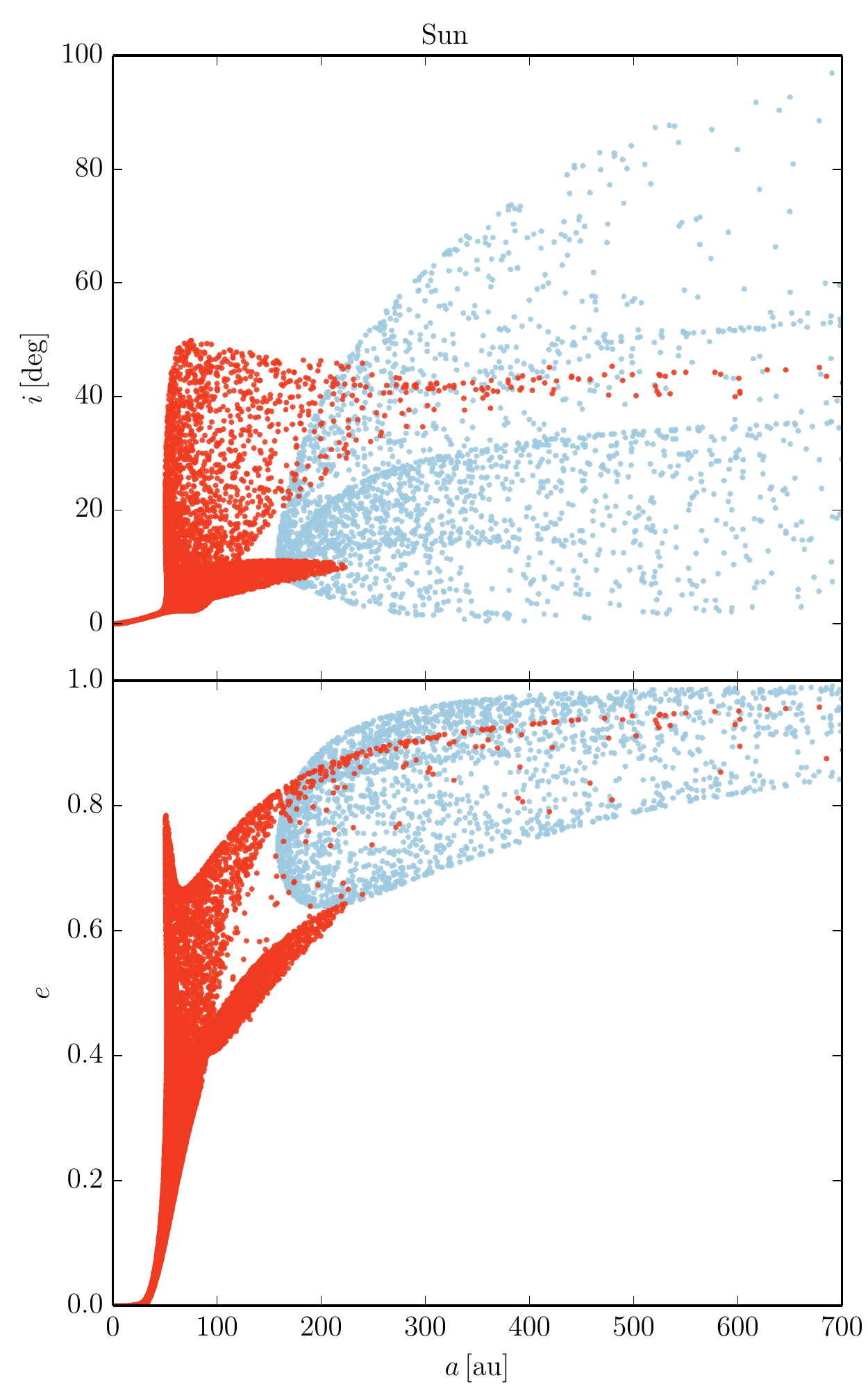}
\includegraphics[width=0.4\textwidth,angle=-0.0]{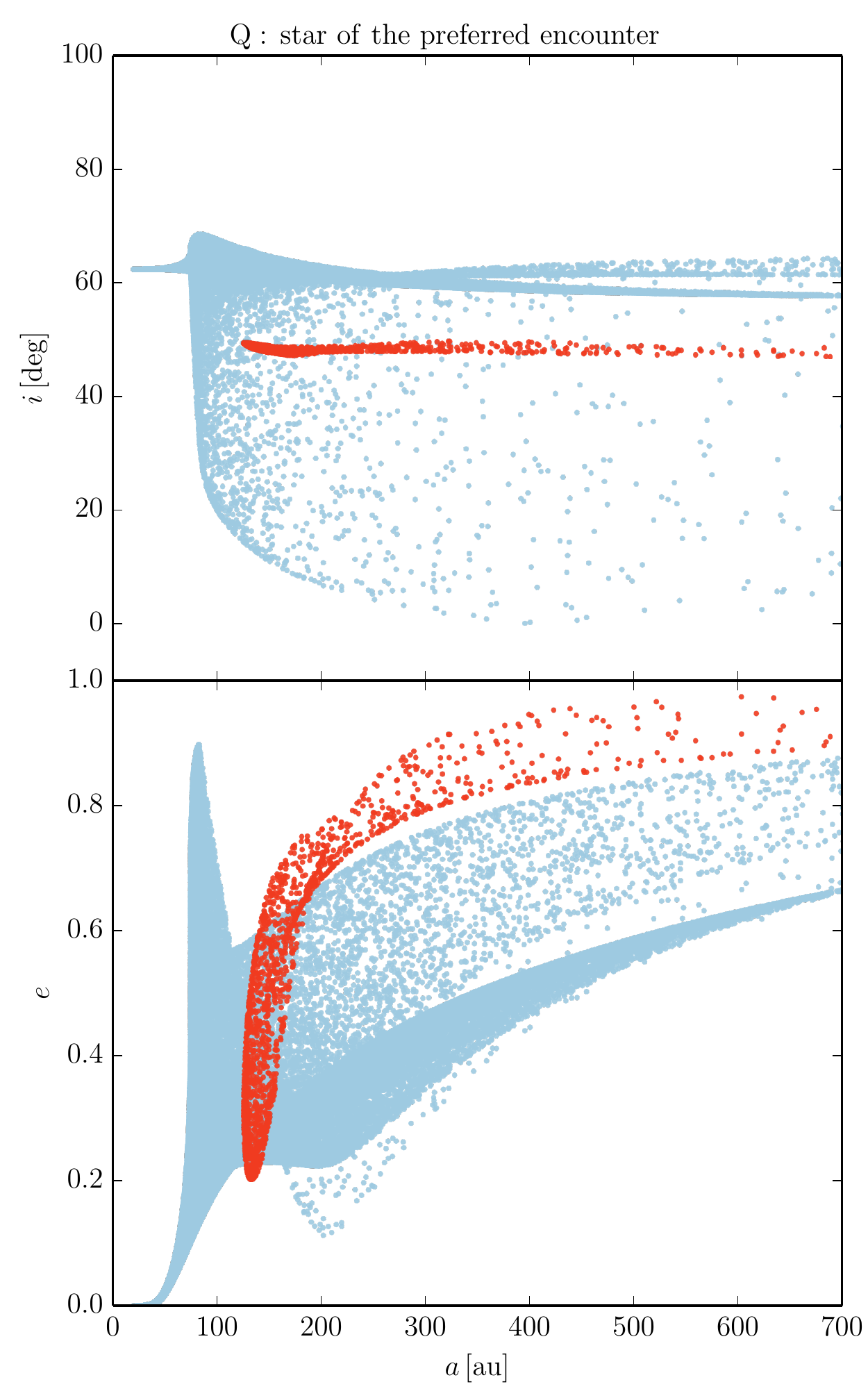}
\end{center}
\caption{Orbital distributions of planetesimals for the Sun (left) and   for the encountering star Q (right) using the preferred encounter parameters (see Tab.\,\ref{Tab:Sednitos}). 
  The top panels give inclination $i$ as a function of semi-major axis $a$, the bottom panels give the orbital eccentricity $e$. 
  The red bullets give the orbital distributions of the planetesimals native of the Sun (assuming its disk extended to 90\,au), the light blue bullets are native to Q. 
  Both initial planetesimal disks are strongly perturbed beyond about 30\,au, but within this distance they are hardly affected \citep[see also][]{2001Icar..153..416K}.
  Note that the simulated particles are not weighted here.
\label{Fig:Encounter}}
\end{figure*}

After constraining the initial conditions that reproduce the Sednitos we rerun the preferred encounter with 100,000 particles in the disk around the encountering star and 100,000 particles around the Sun. 
The disk of the Sun extents from 1\,au to 200\,au.
Some perturbed planetesimals of such disk are members of a native population of Sednitos; we compare this population with the transferred population below.
The results of this calculation are presented in Fig.\,\ref{Fig:Sednitos} and Fig.\,\ref{Fig:Encounter}.
In Fig.\,\ref{Fig:Sednitos}, we compare the observed Sednitos with the captured planetesimals from our best reconstruction of the encounter. 
The orbital distributions of the native and the captured planetesimals are presented in Fig.\,\ref{Fig:Encounter}.

To estimate the number of planetesimals in the captured and the perturbed native population of Sednitos, we adopt a surface density profile $\propto r^{-3/2}$ and a mass of $10^{-3}$\,M$_{\sun}$ for both disks.
We further assume that 10\% of such disks is in the form of Sedna-mass objects (for which we assume $2\cdot10^{21}$\,kg).
In that case the Sun captured a total of $\sim 2600$ planetesimals, 884 of which accreted within the orbit of Neptune (with $q<$30\,au), but most of these are probably ejected by interacting with the planets. 
A total of 936 planetesimals are captured in orbits similar to the observed Sednitos ($q=30$--50\,au or $q=75$--85\,au and $a>150$\,au), and 441 in region between $q=50$\,au and 75\,au. 
The inner Oort Cloud ($q>75$\,au, 150\,au$<a\aplt$1,500\,au) acquired 434 planetesimals, which is consistent with estimates of the current
population of $430^{+400}_{-240}$ \citep{2014Natur.507..471T}.
This would require the planetesimal disk of the encountering star to extend at least to 161\,au, which is a reasonable disk size for a $\sim 1.8$\,$\rm{M}_{\sun}$\, star \citep{2013MNRAS.428.1263B}.

If before the encounter the Sun's disk extended beyond $\sim$90\,au, some of its planetesimals are perturbed to $a$ and $e$ consistent with those of the observed Sednitos.
Assuming the same surface density profile as for Q's disk, 
307 Sun's planetesimals would be perturbed in orbits similar to the observed Sednitos ($q=30$--50\,au or $q=75$--85\,au and $a>150$\,au), 
169 planetesimals would be scattered in the inner Oort Cloud ($q>75$\,au, 150\,au$<a\aplt$1,500\,au), 
and the region between $q=50$\,au and 75\,au would be populated by about 319 native scattered planetesimals.
We use the best encounter parameters for producing the Sednitos from Tab.\,\ref{Tab:Sednitos} to calculate how many of the Sun's planetesimals would be transferred to the encountering star. 
If the solar disk extended to 90\,au, it would have lost $\sim 2.3$\,\% of its planetesimals and $\sim 92$\,\% of those were captured by the other star. 
All the lost planetesimals originate from $a > 70$\,au.
In the right panel of Fig.\,\ref{Fig:Encounter}, we present the distributions of orbital parameters of the Q's own disk particles, and those of the planetesimals it stole from the Solar system. 
These captured objects are in rather curious orbits in the outer parts of the disk around the other star. 
Their inclination is about $14^\circ$ with respect to Q's planetesimal disk and their argument of periastron is clustered around $0^\circ \pm 50^\circ$.

\section{Discussion}
\label{sec:discussion}

During the encounters not only Q's disk is perturbed, but at the same time Q also perturbs the Sun's disk.
In particular, the preferred encounter (Tab.~\ref{Tab:Sednitos}) excites the Sun's disk beyond $\sim 30$\,au\,---\,see left panel of Fig.~\ref{Fig:Encounter}\,---\,in agreement with the disk truncation radius estimate of \citet{2001Icar..153..416K}.
Interestingly, the {\em Nice} model \citep{2005Natur.435..459T,2005Natur.435..462M,2005Natur.435..466G} requires a truncation of the planetesimal disk at $\sim 35$\,au \citep{2004Icar..170..492G}.
Because both values are sufficiently close to be causal and a later subsequent encounter that would truncate the disk at $~35$\,au would also annihilate the population of Sednitos, the capture must have happened before the resonant planetary swap.

The observed Sednitos cluster in the argument of perihelion around $\omega=340^\circ\pm55^\circ$ \citep{2014Natur.507..471T,2014MNRAS.443L..59D}. 
Such clustering is a general characteristic of an exchanged population.
As discussed in Sec.~\ref{sec:omega}, the secular evolution due to the giant planets would cause a precession of $\omega$ on timescales shorter than the age of the Solar system.
If the clustering of $\omega$ is real, 
i.e. it is not a result of an observational bias (see below for more discussion on this issue), a mechanism preserving the distribution of $\omega$ is needed.
The only scenario suggested so far involves a distant planetary-mass object (possibly more than one object) that causes libration of $\omega$ through the Kozai--Lidov mechanism \citep{2006Icar..184..589G,2014Natur.507..471T,2014MNRAS.443L..59D}.
Formation channels for such a super-Earth mass planet were investigated by \citet{2015ApJ...806...42K}, who analyzed three mechanisms: planetary migration from the inner disk, scattering from the inner disk, and in-situ formation.
All three mechanisms require a disk extending up to the orbit of the planet\,---\,disk of gas or small planetesimals circularizes the orbit in the former two scenarios; while the later mechanism requires a reservoir of solid material with a mass $\approx15$\,M$_{\mathrm{Earth}}$ to form a planet at $a\aplt 300$\,au.
However, disk and orbits of any objects beyond $\sim 30$\,au would have been substantially perturbed by the encounter that would deposit Sednitos (see Sec.~\ref{sec:results} and Fig.~\ref{Fig:Encounter}).
Hence, the capture scenario as presented here appears inconsistent with the presence of the outer perturbing planet(s).
The outer companion might have formed later than the Sednitos were transferred, e.g. by a capture of a free floating planet \citep{2012ApJ...750...83P}.
However, in such case it is difficult to explain how the planet acquired the predicted almost circular low-inclination orbit.

Another caveat concerns our assumption that Sednitos are found in the parking zone of the Solar system \citep{2015MNRAS.451..144P}, i.e. that their eccentricities and semi-major axes have been the same since when they acquired their orbits.
However if the outer perturber is present, the Kozai--Lidov oscillations it induces, will also effect the eccentricity and semi-major axis of the Sednitos.
The orbits of the Sednitos could also have been effected by encounters that occurred in the Sun's birth cluster after their delivery.
An encounter as close as we require to deliver the Sednitos (210--320\,au, see Tab.~\ref{Tab:Sednitos}) may have caused the Sun to escape its birth cluster, after which it becomes extremely unlikely to have further close encounters.

It should also be noted that the clustering in $\omega$ might not be a real dynamical feature of Sednitos.
\citet{2014Natur.507..471T} explored possible observational biases and did not identify any that would lead to discovering objects clustered around $\omega=0^\circ$.
Nevertheless, 13 observed objects still provide only small number statistics and therefore we also discuss the possibility that the clustering in $\omega$ is not a real feature of the Sednitos family.
The capture mechanism can still explain the existence of a population of objects in the inner Oort cloud.
The constrains on the population we find are determined by the encounter that is calibrated to deliver the Sednitos assuming that it is a family.
Even if this assumption is wrong, and the 13 objects are not all part of the same family, Sedna and 2012\,VP$_{113}$ can still be explained by a capture.

\section{Conclusions}

The origin of the inner Oort cloud of the Solar system, which is defined as family of planetesimals with $q\apgt 50$\,au and $a\approx 150$--1500\,au \citep{2014Natur.507..471T}, and which currently includes two observed objects\,---\,Sedna \citep{2004ApJ...617..645B} and 2012\,VP$_{113}$ \citep{2014Natur.507..471T}\,---\,is still not well understood.
Here we investigate the scenario where the inner Oort cloud was captured from another star during a close encounter that occurred when both stars, the Sun and its sibling, were still members of their birth cluster \citep{2004AJ....128.2564M,2004Natur.432..598K,2010Sci...329..187L}.
We assume that there are 13 extrasolar objects currently observed in the outer Solar system (with $q>30$\,au, $a>150$\,au), which also share similar inclinations and argument of perihelion ($i=10$--$30^\circ$, $\omega=340\pm55^\circ$, \citealp{2014Natur.507..471T,2014MNRAS.443L..59D}), which we call Sednitos.
Assuming that the orbits of Sednitos have not changed since they were acquired, we reconstruct the encounter that lead to their capture.
The population of objects transferred from a planetesimal disk of the other star during the encounter has in general specific distributions of orbital elements around the star to which it was transferred to.
We use this feature of the captured population and we carry out a Markov Chain Monte Carlo search of the parameter space typical for stellar encounters expected in the Sun's birth cluster.
We provide constrains on the encounters that result in a population of the planetesimals transferred to the Solar system that is consistent with the observed objects.

Understanding the origin of Sednitos and testing the theories for an outer planetary-mass object requires additional observations.
The Gaia astrometric mission is expected to discover $\sim 50$ objects in the outer Solar system.
Being a solar sibling \citep{2009ApJ...696L..13P}, the encountering star may also be discovered in the coming years in the Gaia catalogues.
Having been formed in the same molecular cloud, one naively expects that the chemical composition of this star is similar to that of the Sun \citep{2010MNRAS.407..458B}. 
Finding back our own planetesimals in the predicted orbits around this sibling (see Fig.\,\ref{Fig:Encounter}) would expose the accused robber and would put strong constraints on the extend of the Sun's planetesimal disk.
However, by now the other star has probably turned into a $\apgt 0.6$\,$\rm{M}_{\sun}$\, carbon-oxygen white dwarf, which for a $\sim 1.8$\,$\rm{M}_{\sun}$\, star happens within 2\,Gyr. 
In that case, our stolen stones are probably lost to become free floating planetesimals due to the copious mass loss in the post-asymptotic giant branch phase of the host \citep{2011MNRAS.417.2104V}.

\section*{Acknowledgements}

We thank Inti Pelupessy, Tjarda Boekholt, Adrian Hamers, and Jarle Brinchmann for enriching discussions.
We thank the referee, Hal Levison, for his useful comments which helped to improve the manuscript.
We thank the Leiden/ESA Astrophysics Program for Summer Students (LEAPS) for the support that made this work possible. 
This work was supported by the Interuniversity Attraction Poles Programme initiated by the Belgian Science Policy Office (IAP P7/08 CHARM) and by the Netherlands Research Council NWO (grants \#643.200.503, \#639.073.803 and \#614.061.608) and by the Netherlands Research School for Astronomy (NOVA).
The numerical computations were carried out using the Little Green Machine at Leiden University, and the distributed calculations were performed with the grid-version of AMUSE which was developed with support of the NLeSC.

\bibliographystyle{mn2e}
\bibliography{sedna}

\label{lastpage}
\end{document}